\documentclass[12pt]{iopart}
\usepackage{color}
\usepackage{graphicx}
\usepackage{url}
\begin{document}

\title{Coexistence of chiral symmetry and pseudospin symmetry in one nucleus: triplet bands in $^{105}$Ag}

\author{Hui Jia$^{1}$, Bin Qi$^{1*}$, Chen Liu${^1}$ and Shou-Yu Wang${^1}$}

\address{$1$ Shandong Provincial Key Laboratory of Optical Astronomy
and Solar-Terrestrial Environment, School of Space Science and Physics, Institute of Space Sciences, Shandong University, Weihai, 264209, People's
Republic of China}

\ead{$^*$Corresponding author: bqi@sdu.edu.cn}
\vspace{10pt}

\begin{abstract}
The nearly degenerate triplet bands with the $\pi g_{9/2}^{-1}\otimes \nu h_{11/2} (g_{7/2},d_{5/2})$ configuration in $^{105}$Ag are studied via the relativistic mean-field (RMF) theory and the multiparticle plus rotor model (MPRM). The configuration-fixed constrained triaxial RMF calculations exhibit pseudospin symmetry in single particle spectra and triaxial shape coexistence.
The experimental excitation energies and electromagnetic transition probabilities for the triplet bands are well reproduced by the MPRM calculaitons. The first and second lowest energy bands of the triplet bands are interpreted as pseudospin doublet bands, and the second and third bands are interpreted as chiral doublet bands. The chiral doublet bands show the same phase in the $B(M1)/B(E2)$ staggering, while the pseudospin doublet bands hold the opposite phase. Such triplet bands in $^{105}$Ag provide the first example for the coexistence of chiral symmetry and pseudospin symmetry in one single nucleus.
\end{abstract}
\vspace{2pc}

\submitto{\JPG}

\maketitle


\section{Introduction}
Symmetry is a fundamental concept in  physics. As a many-body quantum system, the atomic nucleus carries an abundance of information associated with symmetries and spontaneous symmetry breaking. Thereinto, chiral symmetry and pseudospin symmetry have been extensively discussed in the past decades.
Chirality in atomic nuclei was predicted by Frauendorf and Meng in 1997~\cite{Frauendorf97}. For a
triaxially deformed nuclear core with a few high-$j$ valence
particles and a few high-$j$ valence holes, the three mutually
perpendicular angular momenta can be arranged to form a
chiral geometry. Due to the chiral symmetry breaking, a pair of nearly degenerate $\Delta I = 1$ bands with the same parity, i.e., chiral doublet bands, should be observed in experiment. So far,
more than 40 candidate nuclei have been reported experimentally in the $A$ $\sim$ 80, 100, 130, and 190 mass regions, see e.g., references~\cite{Starosta01,Koike01,Hecht01,Hartley01,Vaman04,Zhu03,Wang06,Grodner06,Tonev06,Ma18,Timar04,Alcantara04, Timar06,Joshi07,Balabanski04,
Lawrie08,Timar07,Masiteng13,Wang11}.
Theoretically, chiral doublet bands have been
described successfully in many models~\cite{Dimitrov00,Olbratowski04,Mukho07,Wang08,Zhang07,Qi09,Qi11,Koike04, Frauendorf01,Jia16,Meng10,Meng16,Tonev07,Brant08,Chen16,Bhat14,
Chen17}. Based on the
triaxial relativistic mean field (RMF) theory,
it has been suggested that multiple chiral doublet bands (M$\chi$D), i.e.,
more than one pair of chiral doublet bands, could exist in one single nucleus~\cite{Meng06,Peng08,YaoJ09,Li11,Qi13,Li18,Qi18}. The
observations of M$\chi$D~\cite{Ayangeakaa13,Tonev14,Kuti14,Lieder14,Rather14,Liu16,Petrache18} represent important confirmation of triaxial shape coexistence and its geometrical interpretation. A novel type of M$\chi$D bands with the same
configuration was reported in $^{103}$Rh, which showed that chiral geometry can be robust against the increase of the intrinsic
excitation energy~\cite{Kuti14}. The observation of octupole correlations between the M$\chi$D bands in
$^{78}$Br indicates that nuclear chirality can be robust against the octupole correlations~\cite{Liu16}.

Pseudospin symmetry in atomic nuclei was introduced in 1969~\cite{Hecht69, Arima69}. By examining the spherical single-particle spectra,it is found that the two single-particle states with quantum numbers $(n, l, j = l + 1/2)$ and $(n-1, l+2, j = l+3/2)$ are nearly degenerate. References~\cite{Hecht69,Arima69} introduced the so-called pseudospin symmetry and defined the pseudospin doublets as $(\tilde{n} = n-1, \tilde{l}=l+1, \tilde{j} = \tilde{l} \pm 1/2)$ to explain this near degeneracy.
In the deformed nuclei, pseudospin symmetry remains an important physical concept~\cite{Bohr82,Beuschel97}.
So far, lots of phenomena in nuclear structure have been interpreted
by pseudospin symmetry, including nuclear superdeformed configurations~\cite{Dudek87,Bahri92}, identical bands~\cite{Nazarewicz90,Zeng91}, quantized alignment~\cite{Stephens90}, magnetic moments and transitions~\cite{Ginocchio99,Neumann00}, and $\gamma$-vibrational states in nuclei~\cite{Jolos12}.
A pair of nearly degenerate doublet bands with the configuration involving pseudospin doublet states have been observed in several nuclei, e.g., $^{108}$Tc~\cite{Xu08}, $^{118}$Sb~\cite{Wang10}, $^{128}$Pr~\cite{Petrache02}, $^{186}$Ir~\cite{Cardona97}, and $^{195}$Pt~\cite{Petkov07}, and suggested as pseudospin doublet bands. Pseudospin symmetry  has been a hot topic in  nuclear physics,
and an overview of these studies and open problems in the field of nuclear pseudospin symmetry is provided in
references~\cite{Ginocchio05,Liang15} and references therein.

It should be noted that the configurations of some candidate chiral doublet bands involve proton or neutron pseudospin doublet states, e.g., $\pi (f_{5/2},p_{3/2}) \otimes\nu g_{9/2}^{-1}$ in $^{78}$Br~\cite{Liu16},
$\pi g_{9/2}^{-1}\otimes \nu h_{11/2}^{1}(g_{7/2},d_{5/2})$ in $^{103,105}$Rh~\cite{Alcantara04,Kuti14} and $^{105,107}$Ag~\cite{Timar07,Qi13}, and $\pi h_{11/2}^{1}(g_{7/2},d_{5/2}) \otimes \nu h_{11/2}$ in $^{133}$Ce~\cite{Ayangeakaa13}. However, the coexistence of chiral and pseudospin symmetry in one single nucleus has not been discussed so far. We notice that the observed three nearly degenerate negative-parity bands with the same $\pi g_{9/2}^{-1}\otimes \nu h_{11/2} (g_{7/2},d_{5/2})$ configuration in $^{105}$Ag ~\cite{Timar07} might provide an opportunity to investigate the coexistence of chiral and pseudospin symmetry.
The second and third lowest energy bands (labeled G and D in reference~\cite{Timar07}) of the triplet bands exhibited the expected properties of chiral doublet bands, i.e., the very small energy difference between the corresponding states and the same phase in the $B(M1)/B(E2)$ staggering.  
While the phase in the $B(M1)/B(E2)$ staggering of the first lowest energy bands (labeled C in reference~\cite{Timar07}) was opposite to those of the bands G and D. 
Therefore, it is highly interesting to study whether such nearly degenerate triplet bands in $^{105}$Ag are associated with the coexistence of chiral and pseudospin symmetry.

Chiral symmetry and pseudospin symmetry have been successfully described by the relativistic
mean-field (RMF) theory~\cite{Meng06,Peng08,YaoJ09,Li11,Qi13,Li18,Qi18,Ginocchio05,Liang15,Ginocchio97,Meng98,Meng99,Zhou03a} and particle rotor model (PRM)~\cite{Wang08,Zhang07,Qi09,Xu08,Petkov07}. The adiabatic and configuration-fixed constrained triaxial RMF theory has been
used to obtain the triaxial deformations and configurations self-consistently. On the other side, the
quantal calculations for the electromagnetic transition are essential to identify a pair of
nearly degenerate $\Delta I$ = 1 bands with the same parity belonging to chiral or pseudospin
doublet bands, which could be handled in particle rotor model. Based on these considerations, we adopt the triaxial RMF theory and  multiparticle plus rotor model (MPRM) to study the triplet bands in $^{105}$Ag and explore the possible coexistence of chiral and pseudospin symmetry.

\section{Result and discussion}

\begin{figure}
\begin{center}
\resizebox{0.8\textwidth}{!}{\includegraphics{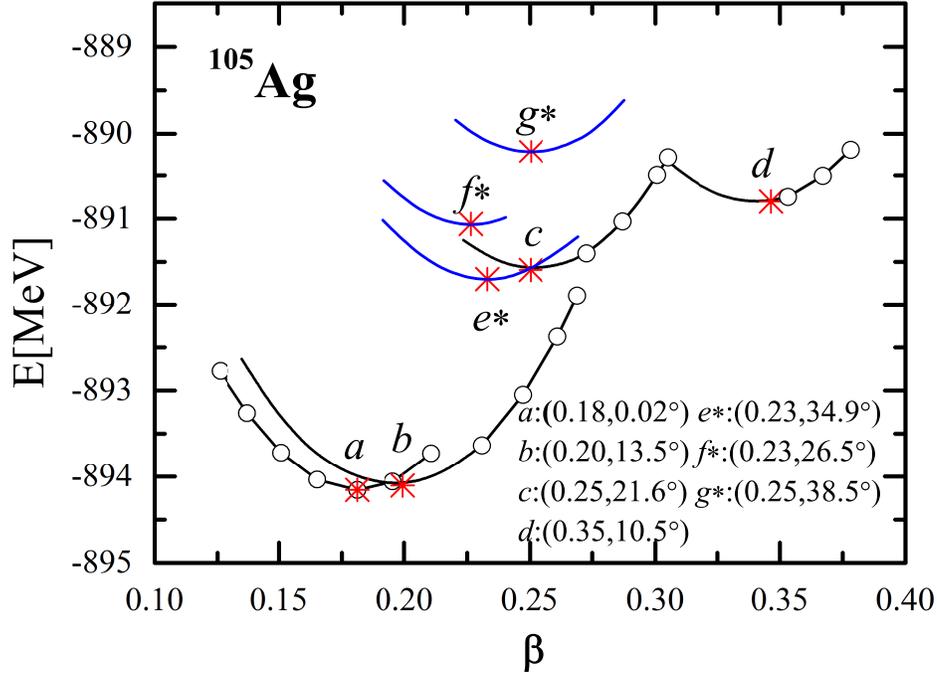}
}
\end{center}
\caption{\label{fig:1}(Color online) The energy surfaces in adiabatic (open
circles) and configuration-fixed (solid lines) constrained triaxial RMF
calculations using effective interaction NL3 for $^{105}$Ag. The minima
in the energy surfaces for the fixed configuration are represented as
stars and labeled as \textit{a, b, c, d, e, f,} and \textit{g}. Their corresponding triaxial
deformation parameters $\beta$ and $\gamma$ are also given. The suitable states for the appearance of the chirality are marked by the blue color and asterisks.
}
\end{figure}

\begin{table}
\caption{\label{Tab:1}The total energies $E_{tot}$, triaxial deformation parameters $\beta$  and $\gamma$ , and their corresponding
 valence nucleon configurations of minima for states \textit{a-g} in the configuration-fixed and $\beta^{2}$-constrained triaxial RMF calculations for $^{105}$Ag, and compared with the experimental excitation energies $E_{x}$. The configuration of the valence nucleons takes reference of 50
nucleons which occupy the states below the 50 major-shell.}
\footnotesize
\begin{tabular}{@{}ccccccc}
\br
 &  \multicolumn{2}{c}{Configuration} &  $E_{tot}$ & &$E_x$(cal.) & $E_x$(exp.) \\
State  & Valence nucleons &  Unpaired nucleons  & (MeV)  & ($\beta, \gamma$) &  (MeV) &(MeV)\\
\mr
$a$  &  $\pi g^{-3}_{9/2}$$\otimes$$\nu(g^{6}_{7/2}d^{2}_{5/2})$ &  $\pi g^{-1}_{9/2}$  &-894.16
& (0.18,0.02)  & 0 & \\
$b$  & $\pi (g^{-2}_{9/2}p^{-1}_{1/2})$$\otimes$$\nu(g^{6}_{7/2}d^{2}_{5/2})$   & $\pi p^{-1}_{1/2}$  & -894.10&(0.20,13.5) & 0.06 & 0 \\
$c$  & $\pi  (g^{-2}_{9/2}g^{1}_{7/2}p^{-2}_{1/2})$$\otimes$$\nu(g^{6}_{7/2}d^{2}_{5/2})$  & $\pi g^{1}_{7/2}$  &-891.60 &(0.25,21.6)   \\
$d$ &$\pi  (g^{-3}_{9/2}g^{2}_{7/2}p^{-2}_{1/2})$$\otimes$$\nu(g^{4}_{7/2}d^{2}_{5/2}h^{2}_{11/2})$ & $\pi g^{-1}_{9/2}$  & -890.81& (0.35,10.5)    \\
$e$$\ast$ &$\pi g^{-1}_{9/2}p^{-2}_{1/2}$$\otimes$$\nu(g^{5}_{7/2}d^{2}_{5/2}h^{1}_{11/2})$  & $\pi g^{-1}_{9/2}$$\otimes$$\nu    h^{1}_{11/2}g^{1}_{7/2}$ & -891.70&(0.23,34.9) & 2.46 & 2.47$\star$ \\
$f$$\ast$ &$\pi  g^{-1}_{9/2}p^{-2}_{1/2}$$\otimes$$\nu(g^{6}_{7/2}d^{1}_{5/2}h^{1}_{11/2})$  & $\pi g^{-1}_{9/2}$$\otimes$$\nu h^{1}_{11/2}d^{1}_{5/2}$ &  -891.06&(0.23,26.5)& 3.10 & 2.62$\dagger$ \\
$g$$\ast$ &$\pi  g^{-1}_{9/2}p^{-2}_{1/2}$$\otimes$$\nu (g^{6}_{7/2}h^{1}_{11/2}h^{1}_{11/2})$  & $\pi g^{-1}_{9/2}$$\otimes$$\nu h^{1}_{11/2}h^{1}_{11/2}$ &  -890.22&(0.25,38.5)& 3.94 & 3.91$\ddagger$\\
\br
\end{tabular}
\begin{indented}
\item[]$\star$ the excitation energy of $I^{\pi}=15/2^{-}$ of band C~\cite{Timar07},
\item[]$\dagger$ the excitation energy of band head $I^{\pi}=15/2^{-}$ of band D~\cite{Timar07},
\item[]$\ddagger$ the excitation energy of band head $I^{\pi}=23/2^{+}$ of band E~\cite{Timar07}.
 \end{indented}
\end{table}

\begin{figure}
\begin{center}
\resizebox{0.8\textwidth}{!}{
  \includegraphics{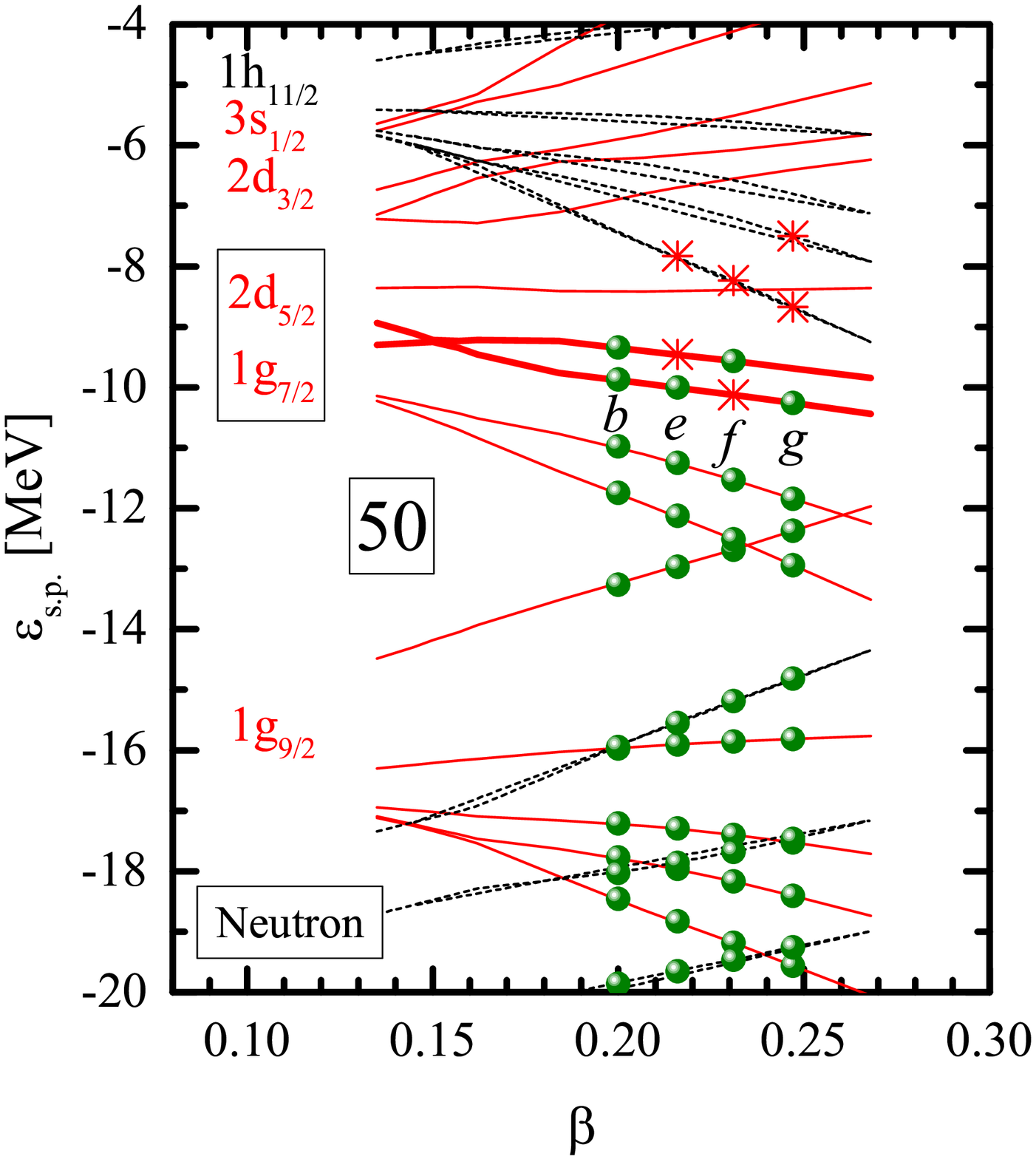}
}
\end{center}
\caption{(Color online) Neutron single-particle levels
obtained in constrained triaxial RMF calculations  as
functions of deformation $\beta$. Positive (negative) parity states are
marked by solid (dashed) lines. Occupations corresponding to the
minima in figure~\ref{fig:1}  are represented by filled circles (two particles) and
stars (one particle).}
\label{fig:2}
\end{figure}

Adiabatic and configuration-fixed constrained triaxial RMF calculations were performed to search for the possible configurations and deformations of $^{105}$Ag. The detailed formulism and numerical techniques of RMF theory can be found in references~\cite{Meng06,Peng08,YaoJ09}. In the present calculations, each Dirac spinor is expanded in terms of a sets of three-dimensional harmonic oscillator bases in Cartesian coordinates with 12 major shells and meson fields with 10 major shells. The effective interaction parameter NL3~\cite{Lala97} is adopted, and the pairing correlation is neglected here. The $\beta^{2}$-constrained calculations have been performed, in which the triaxial deformation  $\gamma$  is automatically obtained by minimizing the
energy.

The energy surface obtained from the adiabatic and configuration-fixed constrained RMF calculations for $^{105}$Ag is shown in figure~\ref{fig:1}. As shown in figure~\ref{fig:1}, several minima observed in the potential energy surfaces are labeled with $a, b, c, d, e, f$, and $g$, in which states $e, f,$ and $g$ are derived from the low-lying particle-hole excitation.
The total energies $E_{tot}$, triaxial deformation parameters $\beta$ and $\gamma$, the corresponding valence nucleon configurations of minima for states $a$$-$$g$, and the experimental excitation energies are listed in table~\ref{Tab:1}. Here, the ground state $a$ with the unpaired nucleon configuration $\pi g_{9/2}$ is obtained from the present RMF calculations. The state $b$ with the unpaired nucleon configuration $\pi p_{1/2}$ corresponds to the observed ground state with the spin and parity $1/2^{-}$~\cite{Rakesh79}. The states $e, f,$ and $g$ are the $\pi g_{9/2}^{-1}\otimes \nu h_{11/2}g_{7/2}$, $\pi g_{9/2}^{-1}\otimes \nu h_{11/2}d_{5/2}$, and $\pi g_{9/2}^{-1}\otimes \nu h^{2}_{11/2}$ configurations, respectively, which correspond to the observed three-quasiparticle bands in $^{105}$Ag~\cite{Timar07}. The excitation energies for states $e, f,$ and $g$ relative to the ground state $a$ are 2.46, 3.10, and 3.94 MeV, respectively, which are in reasonable agreement with the experimental energies of band head $I^{\pi}=15/2^{-}$ for band C (2.47 MeV), that for band head $I^{\pi}=15/2^{-}$ of band D (2.62 MeV) and that of band head $I^{\pi}=23/2^{+}$ for band E (3.91 MeV)~\cite{Timar07}. Furthermore, states $e, f$, and $g$ have the suitable triaxial deformation 34.9$^{\circ}$, 26.5$^{\circ}$ and 38.5$^{\circ}$ together with high-$j$ paticle-hole configurations. Therefore, the chiral doublet bands could be constructed on these states, which might lead to three sets of chiral doublet bands.

Performing the configuration-fixed and $\beta^{2}$-constrained calculations for the state $b$, the neutron single-particle levels as a function of deformation $\beta$ are shown in figure~\ref{fig:2}. The positive (negative) parity states are marked by solid (dashed) lines,
and the occupations corresponding to the minima $b, e, f,$ and $g$ in figure~\ref{fig:1}
are represented by filled circles (two particles) and stars (one particle).
The corresponding quantum numbers for the
spherical case are labeled at the left side of the levels. For positive parity state $g$, the two unpaired neutrons occupy the different 1$h_{11/2}$ levels. For negative parity states, the two unpaired neutrons occupy the $1h_{11/2}$ and $1g_{7/2}$ levels for state $e$, while occupy the $1h_{11/2}$ and $2d_{5/2}$ levels for state $f$.
In fact, it is difficult to distinguish the occupation of the $2d_{5/2}$ and $1g_{7/2}$ orbits, as they mix strongly with each other in the deformed case~\cite{Meng06}.
The levels with $2d_{5/2}$  and $1g_{7/2}$ (denoted by bold lines) accompany each other in a large scale of the $\beta$ region, which manifests the characteristic of the pseudospin symmetry in single particle spectra. These two orbits can be considered as  pseudospin doublet states  $1\tilde{f}_{5/2, 7/2}$. Thus pseudospin doublet bands could be constructed on the states $e$ and $f$.

In order to examine the hypothesis of the chiral doublet bands and pseudospin doublet bands in $^{105}$Ag, quantal MPRM calculations were performed to study the energy spectra and $B(M1)/B(E2)$ ratios for the triplet bands. The detailed formulism of this model can be seen in references~\cite{Qi09,Qi13,Ayangeakaa13}. The $\pi g_{9/2}^{-1}\otimes \nu h_{11/2}g_{7/2}$ configuration with the $(\beta,\gamma)$ = (0.23, 34.9$^{\circ}$) and the $\pi g_{9/2}^{-1}\otimes \nu h_{11/2}d_{5/2}$ configuration with the $(\beta,\gamma)$ = (0.23, 26.5$^{\circ}$) are obtained from the RMF calculations, which present a typical triaxiality. In the present calculations, the same deformation
parameters $(\beta,\gamma)$ = (0.23, 30$^{\circ}$) and moment of inertia $\Im$=32$\hbar^{2}$/MeV are used for the triplet bands. For the electromagnetic
transition, the empirical intrinsic quadrupole moment $Q_{0} =(3/\sqrt{5\pi})R^{2}_{0}Z\beta$, gyromagnetic ratio $g_{R} = Z/A = 0.44$,
$g_{p}(g_{9/2})$ = 1.26, $g_{n}(h_{11/2})$ =$-$0.21, $g_{n}(g_{7/2})$ = 0.26, and $g_{n}(d_{5/2}) $= $-$0.46 are adopted~\cite{Qi11}. Coriolis attenuation factor $\xi$=0.85 for the triplet bands is adopted here.

\begin{figure}[t!]
\begin{center}
\resizebox{0.8\textwidth}{!}{
  \includegraphics{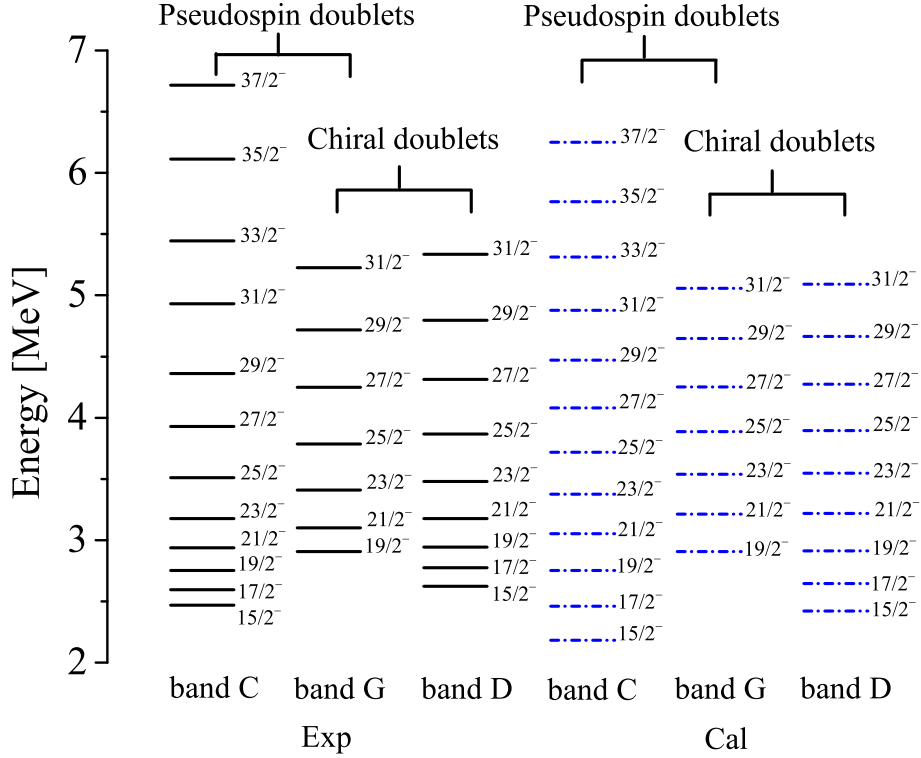}
}
\end{center}
\caption{\label{fig:3} (Color online) Comparison of pseudospin doublets (bands C and G) and chiral doublets (bands G and D)with the corresponding
calculated results using the MPRM model. The $\pi g_{9/2}^{-1}\otimes\nu h_{11/2}g^{-1}_{7/2}$ configuration (for band C) and $\pi g_{9/2}^{-1}\otimes\nu h_{11/2}d^{-1}_{5/2}$ (for bands G and D) with the deformation parameters $(\beta, \gamma)= (0.23,30^{\circ})$ are adopted in the calculations. Coriolis attenuation factor $\xi$=0.85, moment of inertia $\Im$=32$\hbar^{2}$/MeV are used.
}
\end{figure}

The calculated energy spectrum (right panel) in comparison with the experimental data (left panel)~\cite{Timar07} is presented in figure~\ref{fig:3}. The calculated results with the $\pi g_{9/2}^{-1}\otimes \nu h_{11/2} g_{7/2}$ and $\pi g_{9/2}^{-1}\otimes \nu h_{11/2} d_{5/2}$ configurations correspond to the energy spectra of band C and bands G, D, respectively, due to the relative excitation energies of band head and the characteristic of the electromagnetic transitions discussed in the following figure~\ref{fig:4}.
The calculated energy values with the $\pi g_{9/2}^{-1}\otimes \nu h_{11/2} g_{7/2}$ configuration are shifted by +0.679 MeV to coincide with the experimental energy at spin $I$=19/2$\hbar$ of band C. As the observed energy difference between the band heads of band D and band C is 0.15 MeV, the calculated energy values with the $\pi g_{9/2}^{-1}\otimes \nu h_{11/2} d_{5/2}$ configuration are shifted by +0.829 MeV.
It can be seen that the experimental energy spectra, the trend and amplitude of the energy separation between the same spins among the triplet bands are reproduced very well by the MPRM calculations.

\begin{figure}[t!]
\begin{center}
\resizebox{0.8\textwidth}{!}{
  \includegraphics{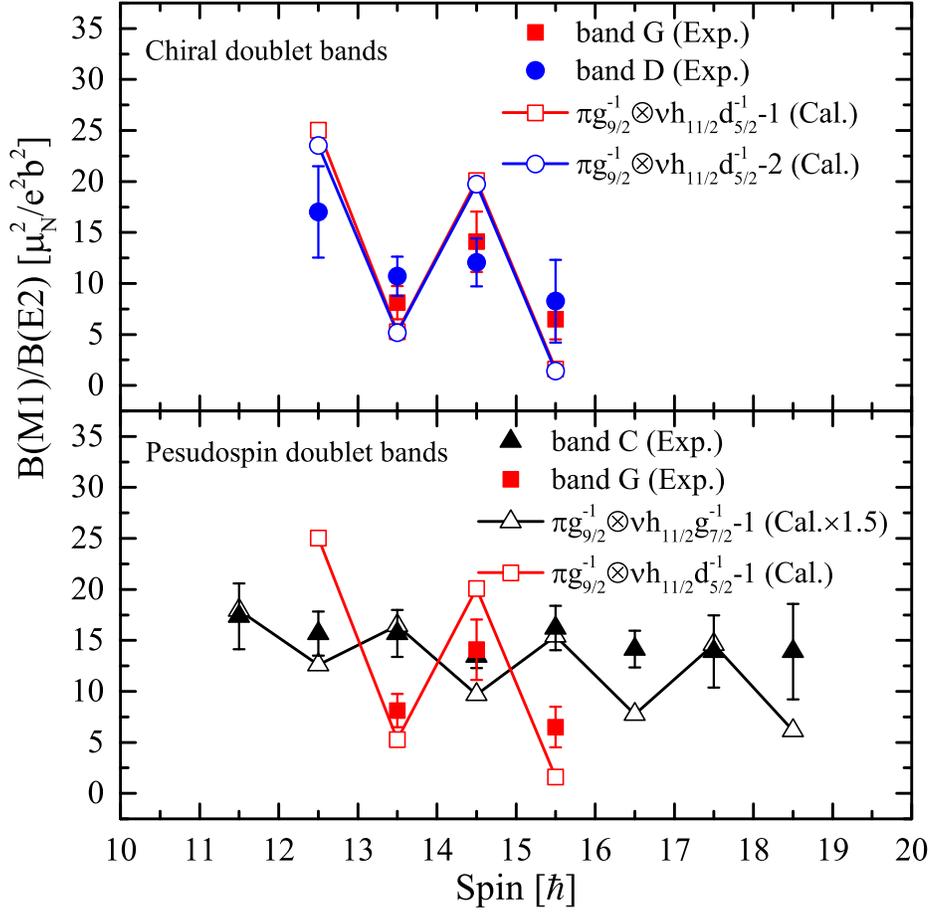}
}
\end{center}
\caption{\label{fig:4} (Color online) The calculated $B(M1)/B(E2)$ values by MPRM as a function of spin, in comparison with the corresponding data available in $^{105}$Ag~\cite{Timar07}.
The same parameters as figure~\ref{fig:3} are adopted.  For the calculations  of electromagnetic
transition, the intrinsic quadrupole moment $Q_{0} =(3/\sqrt{5\pi})R^{2}_{0}Z\beta$, gyromagnetic ratio $g_{R}$ = 0.44,
$g_{p}(g_{9/2})$ = 1.26, $g_{n}(h_{11/2})$ =$-$0.21, $g_{n}(g_{7/2})$ = 0.26, and $g_{n}(d_{5/2}) $= $-$0.46 are adopted~\cite{Qi11}. The number following the configuration label of the theoretical curve corresponds to the energy ordering of the calculated band with the given configuration.
}
\end{figure}
The calculated in-band $B(M1)/B(E2)$ ratios for bands C, G, and D in $^{105}$Ag are presented in figure~\ref{fig:4}, together with the available data extracted from reference~\cite{Timar07}. As shown in the upper panel of figure~\ref{fig:4}, the observed in-band $B(M1)/B(E2)$ ratios in the bands G and D are very similar, and show the same staggering phase, which are consistent with the typical characters for chiral doublet bands~\cite{Vaman04,Koike04}. All these features are well reproduced by the present MPRM calculations with the $\pi g_{9/2}^{-1}\otimes \nu h_{11/2} d_{5/2}$ configuration. Therefore, the present calculations support that the bands G and D with the $\pi g_{9/2}^{-1}\otimes \nu h_{11/2} d_{5/2}$ configuration are chiral doublet bands. As shown in the lower panel of figure~\ref{fig:4}, the observed in-band $B(M1)/B(E2)$ ratios of the band C is reasonably reproduced by the MPRM calculations with the $\pi g_{9/2}^{-1}\otimes \nu h_{11/2} g_{7/2}$ configuration, which show the opposite staggering phase to that of band G. Here to clearly exhibit the staggering phase for band C, the enhancement factor of 1.5 has been used for the calculated values.
The good agreement between the experimental and calculated $B(M1)/B(E2)$ ratios further supports the configuration assignment for bands C and G. Therefore, the present results show that bands C and G indeed belong to pseudospin doublet bands.  Based on the above discussions, bands G, D and bands C, G in $^{105}$Ag can be interpreted as chiral doublet bands and pseudopin doublet bands, respectively.

The present study on the triplet bands of $^{105}$Ag indicates that nuclear chirality can coexist with the pesudospin symmetry. Similar triplet structure has also been observed in neighbouring nucleus $^{105}$Rh (bands 4, 7, 8 in reference~\cite{Alcantara04}), in which the data of electromagnetic transitions for the triplet bands are missing. The measurement
for electromagnetic transition probabilities of $^{105}$Rh in future experiment would be helpful to examine the coexistence of two kinds of symmetries. In fact, when chiral symmetry and pseudospin symmetry occur simultaneously in one single nucleus, two pairs of nearly degenerate $\Delta I=1$  bands with the same parity (``chirality-pseudospin quartet bands'') are expected. It is of highly
scientific interest to search for ``chirality-pseudospin quartet bands'' in $^{105}$Ag and other candidate chiral nuclei.

\section{Summary}

In summary, the nearly degenerate triplet bands with the $\pi g_{9/2}^{-1}\otimes \nu h_{11/2} (g_{7/2},d_{5/2})$ configuration in $^{105}$Ag
are studied via the relativistic mean-field (RMF) theory and the multiparticle plus rotor model (MPRM).
The calculation results show that the first and second lowest energy bands of the triplet bands are pseudospin doublet bands, and the second and third bands are chiral doublet bands.
The triaxial shape coexistence and the pseudospin symmetry in single particle spectra are obtained from the configuration-fixed constrained triaxial RMF calculations. The experimental excitation energies and the electromagnetic transition probabilities for the triplet bands are reproduced very well by the MPRM calculations. The chiral doublet bands show the same phase in the $B(M1)/B(E2)$ staggering, while the pseudospin doublet bands hold the opposite phase. Such triplet bands in $^{105}$Ag provide the first example for the coexistence of chiral symmetry and pseudospin symmetry in one single nucleus.
The present work will motivate the investigation to search for the chirality-pseudospin triplet bands or chirality-pseudospin quartet bands in the nuclei with stable triaxial deformation.

\ack{The authors express sincere thanks to Dr. Q. B. Chen for the discussion of the particle rotor model. This work is partly supported by the National Natural Science Foundation of China (Grants No. 11675094, 11622540, 11705102), the Shandong Natural
Science Foundation (Grants No. JQ201701, ZR2017PA005), the China Postdoctoral Science Foundation (Grant No. 2017M612254) and Young Scholars Program of Shandong University, Weihai (Grant No. 2015WHWLJH01). The computations were carried out on an HP Proliant DL785G6 server hosted by the Institute of Space Science of Shandong University.}

\clearpage


\section*{References}
\begin{thebibliography}{1}
\bibitem{Frauendorf97} S Frauendorf and J Meng, 1997 {\it Nucl. Phys.} A {\bf 617} 131
\bibitem{Starosta01} K Starosta {\it et al.} 2001 {\it Phys. Rev. Lett.} {\bf 86} 971
\bibitem{Koike01} T Koike, K Starosta, C J Chiara, D B Fossan and D R LaFosse 2001 {\it Phys. Rev.} C {\bf 63} 061304(R)
\bibitem{Hecht01} A A Hecht {\it et al.} 2001 {\it Phys. Rev.} C {\bf63} 051302(R)
\bibitem{Hartley01} D J Hartley {\it et al.} 2001 {\it Phys. Rev.} C {\bf64} 031304(R)
\bibitem{Zhu03} S Zhu {\it et al.} 2003 {\it Phys. Rev. Lett.} {\bf91} 132501
\bibitem{Vaman04} C Vaman, D B Fossan, T Koike, K Starosta, I Y Lee and A O Macchiavelli 2004 {\it Phys. Rev. Lett.} {\bf92} 032501
\bibitem{Wang06} S Y Wang, Y Z Liu, T Komatsubara, Y J Ma and Y H Zhang 2006 {\it Phys. Rev.} C {\bf74} 017302
\bibitem{Grodner06} E Grodner {\it et al.} 2006 {\it Phys. Rev. Lett.} {\bf97} 172501
\bibitem{Tonev06} D Tonev {\it et al.} 2006 {\it Phys. Rev. Lett.} {\bf96} 052501
\bibitem{Ma18} K Y Ma {\it et al.} 2018 {\it Phys. Rev.} C {\bf97} 014305
\bibitem{Timar04} J Tim\'{a}r {\it et al.} 2004 {\it Phys. Lett.} B {\bf598} 178
\bibitem{Alcantara04} J A Alc\'{a}ntara-N\'{u}\~{n}ez {\it et al.} 2004 {\it Phys. Rev.} C {\bf69} 024317
\bibitem{Timar06} J Tim\'{a}r, C Vaman, K Starosta, D B Fossan, T Koike, D Sohler, I Y Lee and A O Macchiavelli 2006 {\it Phys. Rev.} C {\bf73} 011301(R)
\bibitem{Joshi07} P Joshi, M P Carpenter, D B Fossan, T Koike, E S Paul, G Rainovski, K Starosta, C Vaman and R Wadsworth 2007 {\it Phys. Rev. Lett.} {\bf98} 102501
\bibitem{Balabanski04} D L Balabanski {\it et al.} 2004 {\it Phys. Rev.} C {\bf70} 044305
\bibitem{Lawrie08} E A Lawrie {\it et al.} 2008 {\it Phys. Rev.} C {\bf78} 021305(R)
\bibitem{Timar07} J Tim\'{a}r {\it et al.} 2007 {\it Phys. Rev.} C {\bf76} 024307
\bibitem{Masiteng13} P L Masiteng {\it et al.} 2013 {\it Phys. Lett.} B {\bf719} 83
\bibitem{Wang11} S Y Wang {\it et al.} 2011 {\it Phys. Lett.} B {\bf703} 40


\bibitem{Dimitrov00} V I Dimitrov, S Frauendorf and F D\"{o}nau 2000 {\it Phys. Rev. Lett.} {\bf84} 5732
\bibitem{Olbratowski04} P Olbratowski, J Dobaczewski, J Dudek and W. Pl\'{o}iennik 2004 {\it Phys. Rev. Lett.} {\bf93} 052501
\bibitem{Mukho07} S Mukhopadhyay {\it et al.} 2007 {\it Phys. Rev. Lett.} {\bf99} 172501
\bibitem{Wang08} S Y Wang, S Q Zhang, B Qi, J Peng, Y M Yao and J Meng 2008 {\it Phys. Rev.} C {\bf77} 034314
\bibitem{Zhang07} S Q Zhang, B Qi, S Y Wang and J Meng 2007 {\it Phys. Rev.} C {\bf75} 044307
\bibitem{Qi09} B Qi, S Q Zhang, J Meng and S Frauendorf 2009 {\it Phys. Lett.} B {\bf675} 175
\bibitem{Qi11} B Qi, S Q Zhang, S Y Wang, J Meng and T Koike 2011 {\it Phys. Rev.} C {\bf83} 034303
\bibitem{Koike04} T Koike, K Starosta and I Hamamoto 2004 {\it Phys. Rev. Lett.} {\bf93} 172502
\bibitem{Frauendorf01} S Frauendorf 2001 {\it Rev. Mod. Phys.} {\bf73} 463
\bibitem{Jia16} H Jia, B Qi, S Y Wang, S Wang and C Liu 2016 {\it Chin. Phys.} C {\bf40} 124103
\bibitem{Meng10} J Meng and S Q Zhang 2010 {\it J. Phys.} G {\bf37} 064025
\bibitem{Meng16} J Meng and P W Zhao 2016 {\it Phys. Scr.} {\bf91} 053008
\bibitem{Tonev07} D Tonev {\it et al.} 2007 {\it Phys. Rev.} C {\bf76} 044313
\bibitem{Brant08} S Brant, D Tonev, G de Angelis and A Ventura 2008 {\it Phys. Rev.} C {\bf78} 034301
\bibitem{Chen16} Q B Chen, S Q Zhang, P W Zhao, R V Jolos and J Meng 2016 {\it Phys. Rev.} C {\bf94} 044301

\bibitem{Bhat14} G H Bhat, R N Ali, J A Sheikh, R Palit 2014 {\it Nucl. Phy.} A {\bf922 } 150
\bibitem{Chen17} F Q Chen, Q B Chen, Y A Luo, J Meng and S Q Zhang 2017 {\it Phys. Rev.} C {\bf96} 051303(R)

\bibitem{Meng06} J Meng, J Peng, S Q Zhang and S G Zhou 2006 {\it Phys. Rev.} C {\bf73} 037303
\bibitem{Peng08} J Peng, H Sagawa, S Q Zhang, J M Yao, Y Zhang and J Meng 2008 {\it Phys. Rev.} C {\bf77} 024309
\bibitem{YaoJ09} J M Yao, B Qi, S Q Zhang, J Peng, S Y Wang and J Meng 2009 {\it Phys. Rev.} C {\bf79} 067302
\bibitem{Li11} J Li, S Q Zhang and J Meng, 2011 {\it Phys. Rev.} C {\bf 83} 037301
\bibitem{Qi13} B Qi, H Jia, N B Zhang, C Liu and S Y Wang 2013 {\it Phys. Rev.} C {\bf88} 027302(R)
\bibitem{Li18} J Li 2018 {\it Phys. Rev.} C {\bf97} 034306
\bibitem{Qi18} B Qi, H Jia, C Liu and S Y Wang 2018 {\it Phys. Rev.} C {\bf98} 014305
\bibitem{Ayangeakaa13} A D Ayangeakaa {\it et al.} 2013 {\it Phys. Rev. Lett.} {\bf110} 172504
\bibitem{Tonev14} D Tonev {\it et al.} 2014 {\it Phys. Rev. Lett.} {\bf112} 052501
\bibitem{Kuti14} I Kuti {\it et al.} 2014 {\it Phys. Rev. Lett.} {\bf113} 032501
\bibitem{Lieder14} E O Lieder {\it et al.} 2014 {\it Phys. Rev. Lett.} {\bf112} 202502
\bibitem{Rather14} N Rather {\it et al.} 2014 {\it Phys. Rev. Lett.} {\bf112} 202503
\bibitem{Liu16} C Liu {\it et al.}, 2016 {\it Phys. Rev. Lett.} {\bf 116} 112501
\bibitem{Petrache18} C M Petrache {\it et al.} 2018 {\it Phys. Rev.} C {\bf97} 041304(R)

\bibitem{Hecht69} K T Hecht and A Adler 1969 {\it Nucl. Phys.} A {\bf 137} 129
\bibitem{Arima69} A Arima, M Harvey and K Shimizu 1969 {\it Phys. Lett.} B {\bf 30} 517
\bibitem{Bohr82} A Bohr, I Hamamoto and B R Mottelson, 1982 {\it Phys. Scr.} {\bf26} 267
\bibitem{Beuschel97} T Beuschel, A L Blokhin and J P Draayer 1997 {\it Nucl. Phys.} A {\bf619} 119
\bibitem{Dudek87} J Dudek, W Nazarewicz, Z Szymanski and G A Leander, 1987 {\it Phys. Rev. Lett.} {\bf59} 1405
\bibitem{Bahri92} C Bahri, J P Draayer and S A Moszkowski 1992 {\it Phys. Rev. Lett.} {\bf68} 2133
\bibitem{Nazarewicz90} W Nazarewicz, P J Twin, P Fallon and J D Garrett 1990 {\it Phys. Rev. Lett.} {\bf64} 1654
\bibitem{Zeng91} J Y Zeng, J Meng, C S Wu, E G Zhao, Z Xing and X Q Chen, 1991 {\it Phys. Rev.} C 44 R1745
\bibitem{Stephens90} F S Stephens {\it et al.} 1990 {\it Phys. Rev. Lett.} {\bf65} 301
\bibitem{Ginocchio99} J N Ginocchio 1999 {\it Phys. Rev.} C {\bf59} 2487
\bibitem{Neumann00} P von Neumann-Cosel and J N Ginocchio 2000 {\it Phys. Rev.} C {\bf62} 014308
\bibitem{Jolos12} R V Jolos, N Y Shirikova and A V Sushkov 2012 {\it Phys. Rev.} C {\bf86} 044320
\bibitem{Xu08} Q Xu {\it et al.} 2008 {\it Phys. Rev.} C {\bf78} 064301
\bibitem{Wang10} S Y Wang {\it et al.} 2010 {\it Phys. Rev.} C {\bf82} 057303
\bibitem{Petrache02} C M Petrache {\it et al.} 2002 {\it Phys. Rev.} C {\bf65} 054324
\bibitem{Cardona97} M A Cardona {\it et al.} 1997 {\it Phys. Rev.} C {\bf55} 144
\bibitem{Petkov07} P Petkov, P vonBrentano, J Jolie and R V Jolos 2007 {\it Phys. Rev.} C {\bf76} 044318


\bibitem{Ginocchio05} J N Ginocchio 2005 {\it Phys. Rep.} {\bf414} 165
\bibitem{Liang15} H Z Liang, J Meng and S G Zhou 2015 {\it Phys. Rep.} {\bf570} 1
\bibitem{Ginocchio97} J N Ginocchio 1997 {\it Phys. Rev. Lett.} {\bf78} 436
\bibitem{Meng98} J Meng, T K Sugawara, S Yamaji, P Ring and A Arima 1998 {\it  Phys. Rev.} C {\bf58} R628
\bibitem{Meng99} J Meng, T K Sugawara, S Yamaji and A. Arima 1999 {\it Phys. Rev.} C {\bf59} 154
\bibitem{Zhou03a} S G Zhou, J Meng and P Ring 2003 {\it Phys. Rev. Lett.} {\bf91} 262501

\bibitem{Lala97} G A Lalazissis, J K\"{o}onig and R Ring 1997 {\it Phys. Rev.} C {\bf55} 540
\bibitem{Rakesh79} R Popli, J A Grau, S I Popik, L E Samuelson, F A Rickey and P C Simm 1979
{\it Phys. Rev.} C {\bf20} 1350

\end{thebibliography}
\end{document}